\documentclass[largeformat]{interact}
\usepackage{epstopdf}
\usepackage[caption=false]{subfig}

\usepackage[numbers,sort&compress,merge]{natbib}
\bibpunct[, ]{[}{]}{,}{n}{,}{,}
\renewcommand\bibfont{\fontsize{10}{12}\selectfont}

\usepackage{braket} 
\usepackage{xcolor} 
\usepackage{blkarray}
\usepackage{multirow}
\usepackage[numbers,sort&compress,merge]{natbib}
\bibpunct[, ]{[}{]}{,}{n}{,}{,}
\renewcommand\bibfont{\fontsize{10}{12}\selectfont}

\usepackage{hyperref}

\begin{document}


\title{Connecting Collisional and Photofragmentation Resonances in the H$_2$ Ungerade Symmetry}

\author{
\name{D\'{a}vid Hvizdo\v{s}\textsuperscript{a}\thanks{CONTACT D. Hvizdo\v{s}. Email: david.hvizdos@univ-lehavre.fr}, Roman \v{C}ur\'{\i}k\textsuperscript{b}, and Chris H. Greene\textsuperscript{c}}
\affil{\textsuperscript{a}Laboratoire Ondes et Milieux Complexes, UMR-6294 CNRS and Université du Havre,
53 Rue de Prony, 76600 Le Havre, France, France\\ \textsuperscript{b}J. Heyrovsk\'{y} Institute of Physical Chemistry, ASCR, Dolej\v{s}kova 3, 18223 Prague, Czech Republic\\ \textsuperscript{c}Department of Physics and Astronomy and Purdue Quantum Science and Engineering Institute, Purdue University, West Lafayette, Indiana 47907 USA}
}

\maketitle

\begin{abstract}
  A recently developed energy-dependent frame transformation theory that incorporates both ionization and dissociation channels of the H$_2$ molecule, is extended to treat the ungerade states that occur both in dissociative recombination and as the final state in ground state photoabsorption.  The theoretical treatment includes the rotational degrees of freedom and is benchmarked against a two-dimensional model that can be solved with high accuracy and also compared with photoabsorption experiments.  Analysis of the resulting spectra shows how the same resonances appear in very different observables, often with quite different line shapes.
\end{abstract}



\section{Introduction}
Our understanding of H$_2$, the most fundamental neutral molecule in nature, has been dramatically improved through a revolution in the theoretical framework that was initiated by Ugo Fano, Ed Chang, Christian Jungen, and Dan Dill,\cite{Fano_PRA_1970,Chang_Fano_1972,Jungen_Dill_JCP_1980} and then pushed through with extensive improvements in the decades since 1980 by Jungen and his collaborators.\cite{Jungen_Dill_JCP_1980,Jungen_Atabek_JCP_1977, Jungen_Raoult_1981, Jungen_Sigma_g_2004, Ross_Jungen_PRL_1987, Sprecher_Jungen_Merkt_2014jcp}  Some of the initial theory was benchmarked against the beautiful early photoabsorption spectra measured by Herzberg and by Dehmer and Chupka.  But the theory and experiments both made still further dramatic strides, culminating in more than two decades of close collaboration between Christian Jungen and Fr\'{e}d\'{e}ric Merkt in a series of striking studies.\cite{Osterwalder_Merkt_Jungen2004jcp,Liu_Jungen_Merkt_2010jcp,Sprecher_Jungen_Merkt2010jcp,Sprecher_Jungen_Merkt_2011FaradayDisc,Sprecher_Jungen_Merkt_2014jcp,Haase_Beyer_Jungen_Merkt_2015jcp,Sommavilla_Merkt_Zsolt_Jungen_2016jcp} 

In the meantime, there has been a parallel push aimed at applying and extending the theoretical techniques advanced by Jungen and his collaborators to other fundamentally important processes occurring in the H$_2$ molecule and its isotopologues, notably the dissociative recombination process, where an electron collides with the positive molecular ion, and triggers molecular dissociation into separated neutrals or ions.  Promising results have been achieved in the theory of dissociative recombination, but perhaps surprisingly, it has never yet been established that the theory can achieve a comparable level of accuracy as has been demonstrated in experimental photoabsorption studies.\cite{HerzbergJungen,glass2022absorption,Sprecher_Jungen_Merkt2010jcp,holsch2023ionization}  Here we pursue our goal of theoretically obtaining such comparably accurate collision cross sections and resonance properties, through a joint study of both dissociative recombination  and photoabsorption in the near-threshold energy range.

In addition to aiming to improve our understanding of collisional problems in the hydrogen molecule, such as dissociative recombination, associative ionization, resonant ion-pair formation, another motivation for the present line of research is our goal to extend our capabilities to other, more complicated molecular target ions. When faced with a new molecular target ion that is subjected to a colliding electron, essentially every theoretical method starts from a fixed-nuclei calculation of a body-frame electron-ion scattering calculation with frozen nuclei. That type of calculation yields a body-frame phaseshift function $\delta({\cal E},R)$ or equivalently, a quantum defect function $\mu_{\ell\Lambda}({\cal E},R) \equiv \delta/\pi$.  In some molecules or symmetries, such as the gerade symmetry of H$_2$, that body-frame scattering amplitude can have off-diagonal elements as well, but the main point is that, in general, that quantity depends on {\it both} the body-frame energy ${\cal E}$ and on the internuclear distance $R$. That energy dependence was neglected altogether in the early, highly successful studies of Jungen and his collaborators, but it has increasingly been realized that an accurate and fully general theoretical method must confront and understand how to incorporate that body-frame energy dependence in the mapping that will determine the final laboratory frame scattering and photoabsorption information.  

The present article extends a theoretical treatment published recently by us that develops a promising generalization of energy-dependent electron-molecule frame transformation theory (EDFT). That study was benchmarked in detail by comparing with an exactly solvable model of the H$_2$ $^1\Sigma_g$ states, probed through the challenging process of dissociative recombination \cite{hvizdovs2025competing}.  While one major focus of that recent article was to test the ability of the EDFT to quantitatively describe that type of collision process, a major component was also to implement a generalized adaptation of the multichannel quantum defect theory treatment of Jungen and Ross \cite{Jungen_Ross_1997} that enables a unified description of interacting ionization and dissociation channels.  In the present article, we adapt the techniques developed in Ref.\cite{hvizdovs2025competing} to treat both resonant dissociative recombination and photofragmentation processes occurring in the H$_2$ ungerade symmetry.  

It is comparatively rare to examine an energy range that is dominated by a richly resonant spectrum for two such different observables as dissociative recombination and photoabsorption, but we demonstrate that it yields insights into both processes. A reason to do this is because, as we know from standard resonance theories from pioneers such as Fano or Feshbach, resonances will occur at the same energy for different observables and with the same total decay width, even if their strength or line shape differ.  Spectroscopic methods are almost always better able to pin down the energy of resonances to far greater precision than electron scattering experiments.  Therefore, when accurate photoabsorption data exist, as is the case for many symmetries of H$_2$, HD, and D$_2$, it is highly recommended to test the precision of a collision calculation or experiment in a difficult process like dissociative recombination by looking for the collisional resonance energies and comparing with photoabsorption resonances known from experiment and/or highly accurate theory.   One promising case where the same theoretical approach was utilized to compute both photoabsorption and dissociative recombination processes using the same MQDT-based treatment was carried out for the H$_3$ molecule,\cite{kokoouline2004photofragmentation} and the encouraging agreement added confidence that the basic mechanism of dissociative recombination in electron collisions with H$_3^+$ was beginning to be understood.  

Another very practical reason exists for why it can be crucial, for theories addressing some applications that need collision data, to provide spectroscopically accurate resonance information.  A well-known case is the dissociative recombination rate of H$_3^+$, where it was crucial for astronomers to determine the DR branching ratio between the ortho and para nuclear spin symmetries, at very low (meV-scale) energies.  After some calculations were carried out to address this, it became increasingly clear that the low energy DR rates for the two symmetries could have a tremendously different branching ratio, depending on whether the low-principal-quantum number perturber closest to threshold was just below or just above threshold.  And for that problem it was crucial to determine this accurately, ideally at the level of 1 meV or better.  It currently appears as though theory has been able to answer that question for H$_3$, consistent with most but possibly not all of the available experimental information.\cite{Albertsson_2014}  For more recent progress on the problem of H$_3^+$ DR and its isotopologues, see Ref.\cite{znotins2025electron}.

The present goal and focus of this article is to adapt the molecular multichannel quantum defect theory (MQDT) to handle dissociation degrees of freedom and also the energy dependence of the body-frame quantum defect function.  Molecular MQDT, pioneered by Jungen, Dill, Fano, and others, efficiently computes a large coupled-channel scattering matrix through the rovibrational frame transformation. Early applications of that theory described photoionization of H$_2$ remarkably well, especially in the vicinity of rotationally- and vibrationally-autoionizing Rydberg states attached to rovibrational thresholds $E_{v^+,N^+}$.\cite{Jungen_Dill_JCP_1980}  

Our recent article on the {\it gerade} states in Ref.\cite{hvizdovs2025competing}, closely related to the simpler {\it ungerade} problem treated here, directly addressed only the $^1\Sigma_g$ states and did not include rotational degrees of freedom.  Here we generalize that formulation to include the coupling of the $^1\Sigma_u$ and $^1\Pi_u$ degrees of freedom, as is well-known to be described accurately by a rotational frame transformation.\cite{Fano_PRA_1970,Chang_Fano_1972,Jungen_Dill_JCP_1980,du1986quantum,Sommavilla_Merkt_Zsolt_Jungen_2016jcp}  We describe this symmetry of the H$_2$ molecule using only $\ell=1$ (p-waves), neglecting the $f$-wave channels that can become important at larger internuclear distances, far from the ground state minimum.  The present study is also restricted to the electronic singlet states of H$_2$, so the spin degrees of freedom and all fine and hyperfine interactions are ignored throughout. More complete studies, including all the relevant spatial and spin symmetries, were carried out for the deutered cation (HD$^+$) \cite{Tamo_etal_HD_PRA_2011,Motapon_HD_PRA_2014}. 

The rovibrational thresholds will be labeled $E_{v^+,N^+}$, with corresponding vibrational eigenstates $\chi_{v^+,N^+}(R) \equiv \langle R| v^+,N^+ \rangle$ of H$_2^+$.  The angular channel functions are denoted $\Phi^{\ell N^+}_{JM}=\langle \hat{r},\hat{R}| (\ell N^+)JM \rangle$ in Hund's case (d) as in \cite{Chang_Fano_1972}, where symmetry limits the ionic angular momentum quantum number $N^+$ to either even or odd values, for para- or ortho-H$_2$, respectively.  In the standard approximation that was typically implemented in the early applications of the rovibrational frame transformation in the 1970s and 1980s, the fixed-nuclei quantum defect function $\mu_{\ell\Lambda}(R)$ for the two relevant body-frame $p$-wave symmetries ($\Sigma,\Pi$) were taken independent of the body-frame energy $\epsilon$.
In that approximation, the electron-ion scattering matrix $S_{v^+N^+,{{v^+}'{N^+}'}}$ and other related matrices of scattering theory (such as the reaction matrix $K_{v^+N^+,{{v^+}'{N^+}'}}$) are immediately obtained through matrix elements of the quantum defect operator $\hat{\mu}_{\ell\Lambda}(R)$ in the ionic vibrational basis set.  This is then combined with the rotational frame transformation from Hund's case (b), where $\mu_{\ell \Lambda}$ is defined, into Hund's case (d) where the molecular ion's angular momentum $N^+$ is a good channel quantum number.  This gives the smooth, short-range scattering matrix, as in Ref.\cite{du1986quantum},
\begin{equation}
    S_{v^+N^+,{{v^+}'{N^+}'}}= \sum_\Lambda \langle N^+|\Lambda \rangle^{\ell J} \langle v^+N^+ | e^{2 i \pi \mu_{\ell\Lambda}(R) } | {{v^+}'{N^+}'}\rangle \langle \Lambda | {N^+}'\rangle^{\ell J}.
\end{equation}
Once this equation is obtained, with ionization channels abbreviated as $i\equiv{v^+N^+}$ that are partitioned at any final laboratory frame energy E into open ($o$) and closed ($c$) channels, the standard closed-channel elimination formulas of MQDT determine the physical scattering matrix in the open channels only, which yield the many closed-channel Rydberg resonances, e.g.:
\begin{equation}
     S^{phys}_{v^+N^+,{{v^+}'{N^+}'}}= S_{oo}-S_{oc}(S_{cc}-e^{-2 i \pi \nu_c})^{-1} S_{co}
\end{equation}
where $\nu_c$ is a diagonal matrix with the effective quantum numbers in the closed vibrational channels, i.e. in a.u., $\nu_{v^+N^+} = [2(E_{v^+N^+}-E)]^{-1/2}$, again with $E$ representing the desired total energy of the system in the laboratory frame.

Once photoabsorption dipole matrix elements were included, the treatment accurately described photoabsorption, photoionization, and photodissociation processes in H$_2$, D$_2$, and HD, following the important article by Jungen and Dill in 1980, which was restricted to only handling H$_2$ photoionization.\cite{Jungen_Dill_JCP_1980}.  Later, Jungen's efforts resulted in techniques for incorporating the molecular dissociation pathway as well, i.e. enlarging the final computed scattering matrix and incoming-wave final states needed for photoionization, to include dissociation channels $d$ in addition to the ionization channels $i$:

\begin{equation}
  {S}=   \begin{bmatrix}
      S_{ii} & S_{id} \\
      S_{di} & S_{dd} \\
  \end{bmatrix}
\end{equation}


\section{\label{sec-2Dmodel}Numerically solvable 2D model}
Ungerade states of H$_2$ represent an example of a dissociative recombination pathway dominated by the indirect mechanism at lower energies. This is because the lowest direct dissociative state attached to the excited $2p\sigma_u$ cation state provides the direct mechanism well outside the Frank-Condon region. In order to benchmark the accuracy of approximations involved in our theory, we proposed \cite{Hvizdos_etal_2018} a 2D model that can be solved numerically without using any physically motivated approximations, such as the Born-Oppenheimer approximation or, in the present extension, the rotational frame transformation.

Our previous model \cite{Hvizdos_etal_2018} for $^1\Sigma_u^+$ states is extended to include also the $^1\Pi_u$ symmetry. Furthermore, we implement a full model Hamiltonian, including the physics of the rotational frame transformation in a way that can be solved exactly. This model Hamiltonian takes the following form for the system studied here
\begin{equation}
\label{eq-fullmodelH}
\begin{aligned}
    H &=  H_e + H_n + \sum_{\Lambda}|\ell\Lambda J \rangle  V^{\ell\Lambda}(R,r) \langle \ell \Lambda J| \\
    &+ \sum_{N^+} |(\ell N^+)J\rangle \frac{N^+ (N^+ +1)}{2M R^2} \langle (\ell N^+)J|\,,
\end{aligned}
\end{equation}
with
\begin{align}
H_\mathrm{e} =& -\frac{1}{2} \frac{\partial^2}{\partial r^2} - \frac{1}{r} + \frac{l(l+1)}{2r^2}, \\
\label{eq-2D_Hen}
H_\mathrm{n} =& -\frac{1}{2M}\frac{\partial^2}{\partial R^2} + U_{1s\sigma}(R)\,.
\end{align}
Here $R$ represents the nuclear separation, $r$ is the distance of the electron from the molecular center, $M$ is the nuclear reduced mass and $U_{1s\sigma}(R)$ is the cation ground state curve. Since the electronic partial wave $\ell$ is restricted to the $p$-wave in this study, i.e. $\ell=1$, the coupling potential's upper $\ell$ index will be omitted in the following. 

The functional form of the potential $V^{\Lambda=0}(R,r)$ coupling the electronic and nuclear degrees of freedom for the $^1\Sigma_u^+$ states is taken from the previous work \cite{Hvizdos_etal_2018}, with one modified parameter value, as
\begin{equation}
\label{eq-Vcoupl_Sigma}
V^{\Lambda=0}(R,r) = -a_1\left(1-\tanh \frac{a_2-R-a_3 R^4}{7}\right) \tanh^4\left(\frac{R}{a_4}\right)\frac{e^{-r^2/3}}{r}\,.
\end{equation}
In case of the $^1\Pi_u$ states this potential is sought in the form
\begin{equation}
\label{eq-Vcoupl_Pi}
V^{\Lambda=1}(R,r) = e^{-r^2/\gamma^2} \sum_{i=1}^2 a_i e^{-\left[(R-b_i)/c_i\right]^2}.
\end{equation}

\begin{table}[h]
\caption{\label{tab-Vcoupl_Pi} Parameters of the $V^{\Lambda}$ potentials (\ref{eq-Vcoupl_Pi}) are given in atomic units.}
\begin{tabular}{lcccccc}
\hline
\multirow{2}{*}{$^1\Sigma_u^+$} & $a_1$ & $a_2$ & $a_3$ & $a_4$ & & \\
                                & 1.6435 & 6.2000 & 0.0125 & 1.0200 \\              
\hline
\multirow{2}{*}{$^1\Pi_u$} & $a_1$ & $b_1$ & $c_1$ & $a_2$ & $b_2$ & $c_2$ \vspace{2mm}\\
                           & 0.2592 & 3.6464 & 2.2558 & 0.0620 & 5.9278 & 3.1500 \\
\hline
\end{tabular}
\end{table}
The range of the electronic potential is fixed here at $\gamma = 2$ a.u..
Variable parameters present in the Eqs.~(\ref{eq-Vcoupl_Sigma}) and (\ref{eq-Vcoupl_Pi})  were optimized to fit three quantum defect curves obtained by the Rydberg formula from the three lowest adiabatic curves published by \citet{Wolniewicz_Staszewska_Sig_JMS_2003,Wolniewicz_Staszewska_Pi_JMS_2003}. Values of these parameters are given in Tab.~\ref{tab-Vcoupl_Pi}. An average error, averaged over the data points of the three curves, can be estimated from the fitting procedure as $\Delta \mu_\Sigma \sim 1.1 \times 10^{-2}$ and $\Delta \mu_\Pi \sim 2.1 \times 10^{-3}$.

\begin{figure}[tbh]
\begin{center}
\includegraphics[width=0.6\textwidth]{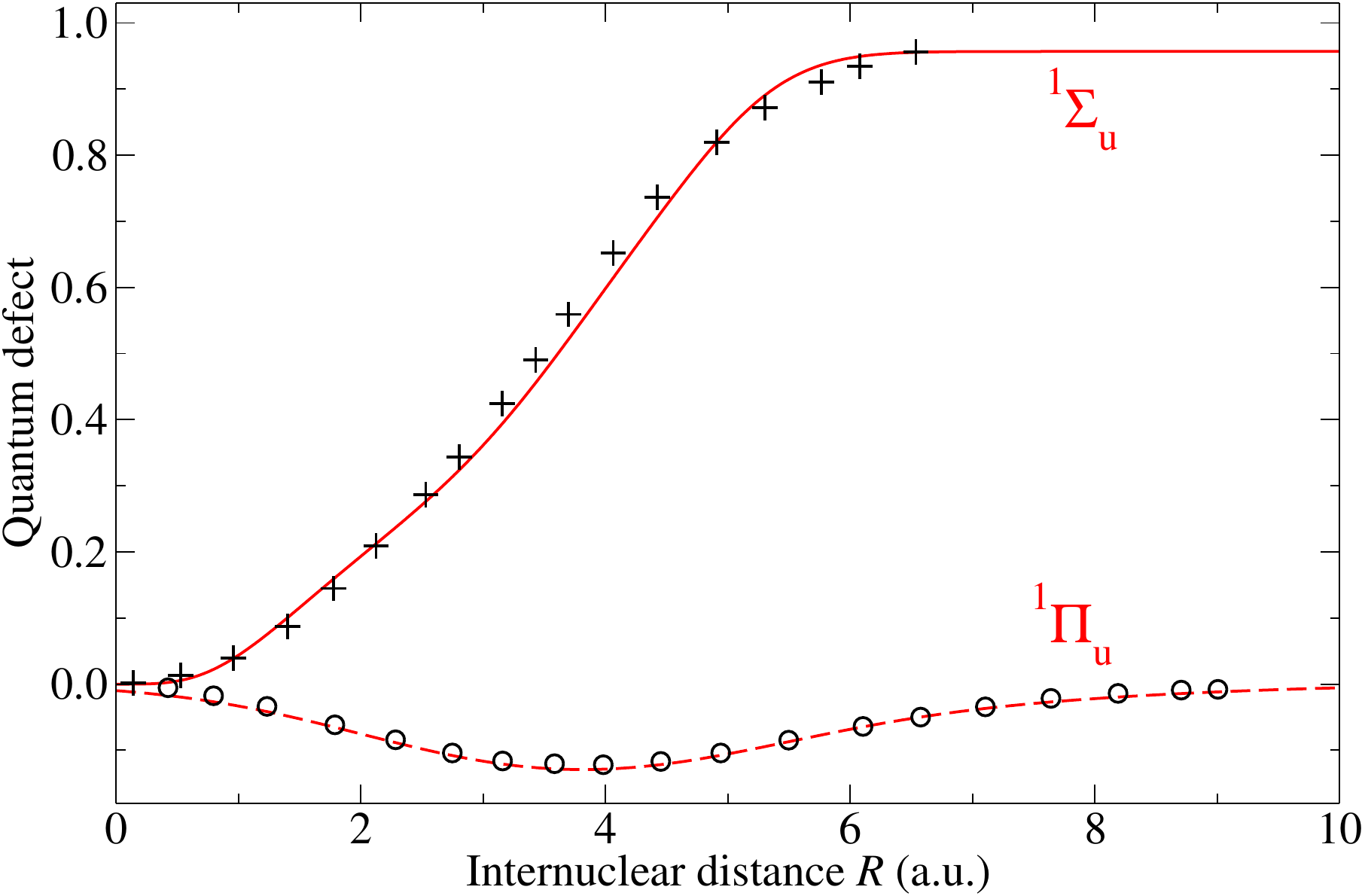}
\caption{\label{fig-Vqd}
Computed fixed-nuclei zero-energy $p$-wave quantum defects for $^1\Sigma_u^+$ (full curve) and $^1\Pi_u$ (broken curve) symmetry compared with the data extracted from \citet{Jungen_Atabek_JCP_1977} shown as crosses ($^1\Sigma_u^+$) and circles ($^1\Pi_u$).
}
\end{center}
\end{figure}

The Hamiltonian (\ref{eq-fullmodelH}) is written in two incompatible sets of quantum numbers. The kets $|(\ell N^+)J\rangle$ span the angular Hilbert space in the laboratory frame, while the body frame is described by $|(\ell\Lambda) J\rangle$.
Transformations between these two representations are controlled by the projections, as formulated by Chang and Fano \cite{Chang_Fano_1972}:
\begin{equation}
   Y_{N^+,\Lambda} \equiv \langle N^+ |\Lambda \rangle ^{(\ell \eta J)}= (-1)^{J+\Lambda-N^+} \langle (\ell J)N^+ 0 | \ell -\Lambda, J \Lambda \rangle \frac{1 + \eta (-1)^{J-\ell-N^+}}{\sqrt{2(1+\delta_{\Lambda 0})}}
\end{equation}
The treatment in the present article will be restricted to the final $p$-wave electron states having $\eta=1$ in the field of even parity para-states of H$_2^+$ with even values of $N^+$.  We ignore the hyperfine and other magnetic interactions, and assume the two electrons are coupled into a singlet spin state throughout.
The nonzero row indices are $N^+ = 0,2$ and the column indices are $\Lambda = 0,1$, respectively, giving a 2$\times$2 unitary matrix for the transformation coefficients $\langle N^+ |\Lambda \rangle ^{(\ell \eta J)}$:
\begin{equation}
\label{eq-LambdaN_matrix}
\begin{blockarray}{llcc}
 & \Lambda & 0 & 1 \\
 N^+ & & & \\
 \begin{block}{ll(cc)}
 0 & & \frac{1}{\sqrt{3}} & \sqrt{\frac{2}{3}} \\
 2 & & -\sqrt{\frac{2}{3}} & \frac{1}{\sqrt{3}} \\
 \end{block}
\end{blockarray}
\end{equation}

\subsection{\label{ssec-2Dexact-details}Details of the numerically exact solution}

Scattering eigenstates of the 2D model Hamiltonian (\ref{eq-fullmodelH}) are computed in the laboratory frame where the coupling between the $N^+=0$ and $N^+=2$ channels is provided by its third term.
The size of the 2D $R$ matrix box was defined by nuclear and electronic boundaries, $R_0 = 12$ a.u. and $r_1 = 50$ a.u., respectively. The nuclear coordinate $R$ was represented by 300 B-splines while the electronic coordinate $r$ used 100 B-splines. The $R$ matrix formed on the surface employed 14 ionization and 5 dissociation channels. The stability of the results was checked for several box sizes, namely $R_0 = 8,10,12,15$ a.u. and $r_1=50,60,80$ a.u.

The 2D model provides (see Eq.~(26) in Ref.~\cite{Curik_HG_2DRmat_2018}) the $R$ matrix $R$ on both ionization and dissociation surfaces. The channels spanning the ionization (electron fragmentation) surface are described by quantum numbers $(v^+ \ell N^+)$. The channels $(d)$ spanning the dissociation surface (nuclear fragmentation) are obtained by diagonalization of the $r$-dependent parts of the Hamiltonian (\ref{eq-fullmodelH}) on the dissociation surface. While these channels have nonzero contributions in both laboratory frame channels with $N^+=0,2$, they are exactly diagonal in the $\Lambda$ channels which are obtained through the rotation by the unitary transformation (\ref{eq-LambdaN_matrix}). This is because our model Hamiltonian (\ref{eq-fullmodelH}) can be rotated exactly to the diagonal $\Lambda$ representation on the dissociation surface since it does not contain any explicit non-Born-Oppenheimer rotational terms that would couple electronic and rotational angular momenta on the dissociation surface. Note that the $\Lambda$-diagonal dissociation channels result here from the form (\ref{eq-fullmodelH}) of our model Hamiltonian. However, this property was also generally accepted in previous DR studies \cite{Takagi_H2rot_JPB_1993,Schneider_JPB_1997}.

The $K$ matrix is obtained by the well-known expression \cite{Aymar_Greene_LKoenig_1996}
\begin{equation}
\label{eq-R2K}
K^{J} = \left(F^{J} -F'^{J} R^J \right)
\left(G^{J} -G'^{J}R^J \right)^{-1},
\end{equation}
where $F^{J}$ and $G^{J}$ are diagonal matrices containing regular and irregular solutions on both surfaces, respectively. On the electronic surface they are the regular and irregular Coulomb functions $f_\ell, g_\ell$ corresponding to the asymptotic centrifugal term $l(l+1)/2r^2$, while on the nuclear surface these should be regular and irregular spherical Riccati-Bessel functions $F^{J}_d, G^{J}_d$ corresponding to the centrifugal term
\begin{equation}
\label{eq-centrifugal_J}
\frac{J(J+1) - \Lambda^2}{2MR^2}.
\end{equation}
The expression above is a result of non-trivial physics of the angular momenta projections between the laboratory and body frames, physics that is not included correctly in our model. As mentioned above, the Hamiltonian (\ref{eq-fullmodelH}) can be rotated exactly to the diagonal $\Lambda$ representation. When the $N^+(N^++1)$ centrifugal potentials are rotated as well, they result in the effective centrifugal potential in the $\Lambda$ channels
\begin{equation}
    \label{eq-model_barrier}
    \frac{\sum_{N^+} \langle \Lambda|N^+ \rangle N^+(N^++1) \langle N^+ | \Lambda \rangle}{2MR^2}\,,
\end{equation}
where the superscript ${(\ell \eta J)}$ was omitted in the transformation matrices. These barriers are twice stronger than those of Eq.~(\ref{eq-centrifugal_J}). 
Both centrifugal potentials (\ref{eq-centrifugal_J}) and (\ref{eq-model_barrier}) yield spherical Riccati-Bessel functions with a non-integer order index. However, apart from the narrow energy windows at the dissociation thresholds, an energy range not considered here, the integral cross sections do not depend on the order of the matching Bessel functions.

\section{\label{sec-rotft}Rovibrational frame transformation}

The rotational frame transformation (FT) \cite{Chang_Fano_1972} is an approximate procedure developed to connect the body frame rotational quantum numbers $(\ell,\Lambda)$ to the laboratory frame quantum numbers $(\ell,N^+)$ on the electronic fragmentation surface. However, on the nuclear fragmentation surface the electronic angular momentum $\ell$ is already absorbed into the system and thus the total angular momentum $\vec{J}=\vec{N}^++\vec{\ell}$ should be used in both frames.  The main point of the rotational frame transformation concept is that there are two incompatible quantum operators with corresponding locally good quantum numbers.  One is the Hund's case (b) quantum number $\Lambda$ that is appropriate at short electron-ion distances where the outermost electron has relatively high kinetic energy, and the other is $N^+$, the molecular ion angular momentum quantum number, appropriate in Hund's case (d), relevant when the outermost electron is farther away and its kinetic energy depends on which rovibrational energy level the molecular ion resides in.

\subsection{Calculation steps to treat competing ionization and dissociation in the energy-dependent FT}

The 2D model described in the preceding section can be solved in two different ways.  One way, discussed in subsection \ref{ssec-2Dexact-details}, uses the 2D R-matrix method inside the volume $r < r_0$ and $R<R_0$, treating fully and essentially exactly the coupling between the electronic and nuclear degrees of freedom, followed by matching to Coulomb functions in the electron coordinate $r$ and dissociative (Bessel) solutions in the dissociative coordinate $R$. But the point of the present section is to implement the energy-dependent frame transformation (EDFT) theory, to assess its validity, accuracy, and limitations. The EDFT results can then be benchmarked against the accurate 2D R-matrix results.

\subsubsection{Fixed-nuclei quantum defect and transition-dipole functions}
The very first step in implementing the EDFT is to solve the fixed-$R$ Schr\"{o}edinger equations for the $p\sigma$ and $p\pi$ symmetries, which produces the quantum defect functions $\mu_\Sigma({\cal E},R)$ and $\mu_\Pi({\cal E},R)$.  As is usual in MQDT studies, $\pi \mu_\Lambda$ can be regarded as the phaseshift relative to the ``energy-normalized'' Coulomb functions $f_\ell({\cal E},r),g_\ell({\cal E},r)$.  This means that for ${\cal E}<0$, the finiteness boundary condition at $r\rightarrow \infty$ is not yet imposed, although of course it will ultimately be enforced in the usual ``closed-channel elimination step'' of MQDT.  The body-frame ${\cal E}=0$ energy coincides at each $R$ value with the energy of the $1s\sigma$ potential energy curve of H$_2^+$.  In the present study, these quantum defect functions are tabulated on a grid of ${\cal E}$ and $R$ and later interpolated because they will be integrated over $R$ in subsequent steps of the calculation.

\begin{figure}[tbh]
\begin{center}
\includegraphics[width=0.6\textwidth]{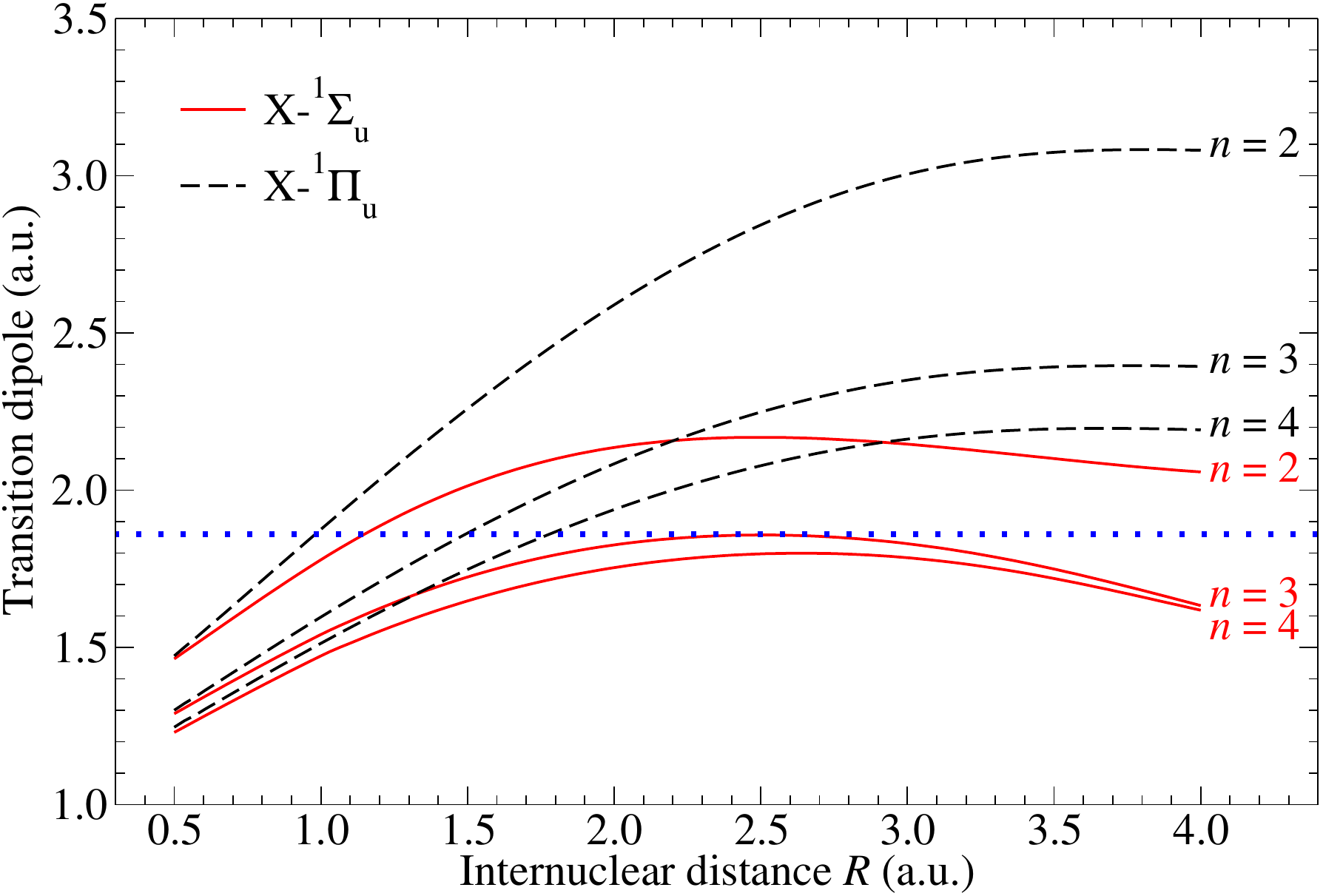}
\caption{\label{fig-dipole}
Energy-normalized transition dipole moments between the ground state X$^1\Sigma_g^+$ and the three lowest states in the $^1\Sigma_u^+$ and $^1\Pi_u$ symmetries. The horizontal dotted line displays the constant value of 1.86 a.u. used in the Ref.~\cite{Jungen_Dill_JCP_1980} for both symmetries.
}
\end{center}
\end{figure}
When electric dipole photoabsorption from the molecular ground state is of interest, dipole matrix elements are also needed between the unity-normalized ground state and the body-frame energy-normalized final states.  In the present study, the initial state will be the X $^1\Sigma_g^+$ state of para-H$_2$.  The $2D$ model for the ungerade states does not describe that ground state, of course, so the dipole elements are adapted from the accurate {\it ab initio} calculations by \citet{Wolniewicz_Staszewska_Sig_JMS_2003,Wolniewicz_Staszewska_Pi_JMS_2003} . Those electric dipole matrix elements will be denoted $d_\Lambda({\cal E},R)$ and connect the molecular ground state to the final ungerade states of H$_2$ that have the quantum defect functions $\mu_\Lambda({\cal E},R)$.  

The {\it ab initio} dipole elements \cite{Wolniewicz_Staszewska_Sig_JMS_2003,Wolniewicz_Staszewska_Pi_JMS_2003} computed for $n=2,3,4$ unity-normalized wave functions were renormalized by the $\nu^{3/2}$ factor, which makes them ``energy-normalized'' in the MQDT sense. Their $R$-dependence is shown in Fig.~\ref{fig-dipole} together with the constant value of 1.86 a.u. used in the study of \citet{Jungen_Dill_JCP_1980}. The energy dependence of the dipole function is generated here from the three data points by linear interpolation or extrapolation from the central $n=3$ point through the $n=2$ and $n=4$ points to the lower and higher energies, respectively.

\subsubsection{Energy-dependent laboratory frame reaction matrix calculation}
We use an accurate ground state potential energy curve for the $1s\sigma$ state of the molecular ion to compute the vibrational wavefunctions $\chi^{(b)}_{v^+,N^+}(R)$ of H$_2^+$ in a finite range of internuclear distances, $0 < R < R_0$.  The present calculations adopt the value $R_0=12$ a.u. and $r_0=8$ a.u. As was discussed in our related treatment of the EDFT for the gerade states of H$_2$ in Ref.\cite{hvizdovs2025competing}, two different sets of orthonormal vibrational eigensolutions are needed.  The two sets of wavefunctions each obey a vanishing boundary condition at the origin but different logarithmic derivatives ($\chi^{(b)'}(R_0)/\chi^{(b)}(R_0)=-b$) at $R=R_0$.  The first set has $b \rightarrow \infty$, corresponding to a vanishing wavefunction boundary condition, while the second set has $b=0$, corresponding to vanishing first derivative at $R_0$. Because the present study considers only the final state with $J=1,\eta=1,\ell=1$, there are two states $\chi^{(b)}_{v^+N^+}(R)$ with $N^+=0$ and $2$ that are computed for each vibrational state $v^+$ having $v^+$ nodes.  The solutions obey the following differential equation in the ionic potential curve for H$_2^+$, which includes the diagonal adiabatic correction\cite{Peek_Madsen_1971}:
\begin{equation}
    -\frac{1}{2M} \frac{d^2 \chi^{(b)}_{v^+N^+}(R)}{dR^2} + \left(\frac{N^+(N^++1)}{2 M R^2} + U_{1s\sigma}(R) - E^{(b)}_{v^+N^+} \right)\chi^{(b)}_{v^+N^+}(R)=0.
\end{equation}
One way to understand why more than one value of the surface logarithmic derivative $-b$ are needed is because the high-$v^+$ wavefunctions need to be able to represent a dissociative radial wavefunction near the boundary $R_0$.  If only ionic vibrational solutions with one value of $b$ were utilized, these would only be able to describe internuclear dissociating wavefunctions with that same value of $b$.

Now, following the logic laid out in Sec.III of Ref.\cite{hvizdovs2025competing}, for each $b$-value on the dissociative surface, a set of linearly-independent frame-transformed solutions at energy $E$ and angular momentum $J$ in the region outside the reaction volume are computed by the frame transformation procedure, which involves a numerical quadrature.  In the first step, in the spirit of Refs.\cite{Gao_Greene_PRAR_1990, EDFT_HeHplus_2020,Backprop_FT_2020,hvizdovs2025competing}, we write a wavefunction valid at short distances $r \sim 10$ a.u. whose vibrational wavefunction is a specific ionic eigenstate for logarithmic derivative $-b$, i.e., transformed now into Hund's case (d).  The solution for channel $i' \equiv \{{v^+}',{N^+}'\}$ is equal to:
\begin{equation}
\label{eq-first-dipole}
\begin{aligned}
    &\Psi_{i'}^{(b)} =\\
    &\sum_\Lambda |\ell \Lambda\rangle^{(J)} \biggl( f(r) \cos{ \pi \mu_\Lambda({\cal E}_{i'},R)} - g(r) \sin{ \pi \mu_\Lambda({\cal E}_{i'},R)} \biggr) \langle \Lambda|{N^+}'\rangle^{(\ell \eta J)} \chi^{(b)}_{{v^+}'{N^+}'}(R) 
\end{aligned}
\end{equation}
Observe that the energy dependence of the $p$-wave Coulomb functions has not been indicated in the preceding equation, because they are nearly energy-independent in the small-$r$ region, an assumption that is part of the logic of the rovibrational frame transformation theory.  But next comes the crucial realization that as this solution evolves into the outer region $r \gtrsim 10$ a.u., and with the unit operator applied from the left in the form $\hat 1 = \sum_{v^+N^+} |\Phi_i \rangle \langle \Phi_i |$, the following form of the  $i'$-th laboratory frame solution is seen to be:
\begin{equation}
\label{eq-EDFT_psi_out}
    \Psi^{(b)}_{i'}(E,r,R) = \sum_{i} 
    \Phi^{(b)}_i
    \bigg(f(\epsilon_{i},\ell_{i},r) \mathcal{C}^{(b)}_{i i'}(E) 
    - g(\epsilon_{i},\ell_{i},r) \mathcal{S}^{(b)}_{i i'}(E) \bigg),
\end{equation}
at $r>r_0$.
Here the electronic channel functions are defined by 
\begin{equation}
    \Phi^{(b)}_i \equiv \ket{(\ell N^+_i)J}\chi^{(b)}_{v^+_i,N_i^+}(R),
\end{equation}
and the frame-transformed representation of the scattering information is contained in the two matrices:
\begin{equation}
\label{eq-EDFT_match}
\begin{aligned}
    \mathcal{S}^{(b)}_{ii'}(E) &= \sum_\Lambda \langle N^+_i | \Lambda \rangle \biggr( \int_{0}^{R_0} \chi_{i}(R) \sin{\pi \mu_\Lambda(\mathcal{E}_{i'},R) }\chi_{i'}(R) \text{d}R \biggr ) \langle \Lambda|N^+_{i'} \rangle , \\
    \mathcal{C}^{(b)}_{ii'}(E) &=
    \sum_\Lambda \langle N^+_i | \Lambda \rangle \biggr( \int_{0}^{R_0} \chi_{i}(R) \cos{\pi \mu_\Lambda(\mathcal{E}_{i'},R) }\chi_{i'}(R) \text{d}R \biggr) \langle \Lambda|N^+_{i'} \rangle .
\end{aligned}
\end{equation}
In these integrals, the choice of the body-frame energy ${\cal E}_{i'}$ is crucial; following the logic of Ref. \cite{Gao_Greene_PRAR_1990,Backprop_FT_2020,hvizdovs2025competing}, ${\cal E}_{i'}$ is chosen at energy $E$ to equal the electronic channel energy in rovibrational channel $i'$, i.e. ${\cal E}_{i'}=E-E_{i'}$. In Eq.~(\ref{eq-EDFT_match}), for brevity, we simplify the notation for the rotational frame transformation element introduced in the previous section to $\langle N^+_i | \Lambda \rangle \equiv \langle N^+_i |\Lambda \rangle ^{(\ell \eta J)}$.  

For photoabsorption processes from the ground state, the EDFT procedure above must similarly be applied to obtain the reduced dipole matrix element from the ground state in the initial potential curve, with vibrational state $v_0$ and initial angular momentum (taken here to be $J_0=0$), to the final $J=1$ laboratory frame state shown above in Eq.~(\ref{eq-EDFT_psi_out}), denoted $\Psi^{(b)}_{i'}(E,r,R)$, (see Eq. (10), Ref.\cite{du1986quantum}):
\begin{equation}
\label{FT-dipolematrix}
\begin{aligned}
    &d^{(b)}_{i'}\equiv d^{(b)}_{{v^+}',{N^+}'} = \\
    &\sum_\Lambda \langle  {N^+}'|\Lambda \rangle^{(\ell \eta J)} \biggl( \int_0^{R_0} \chi^{(b)}_{{v^+}',{N^+}'}(R)  d_\Lambda({\cal E}_{{v^+}'{N^+}'},R)   \chi^{\rm X}_{v_0,J_0}(R) dR \biggr)  \langle \Lambda | J_0 \rangle^{(1,J)} \sqrt{2J+1}.
\end{aligned}
\end{equation}
Note that the body-frame dipole matrix elements $d_\Lambda({\cal E},R)$ are the matrix elements from the ground state to the final energy-normalized body frame states having the asymptotic form written in the large parentheses of Eq.~(\ref{eq-first-dipole}).
After these ${\cal C,S}$ matrices have been formed, the reaction matrix $K$ can be computed for each $b$ as ${\cal S C}^{-1}$. The transition dipole above, viewed as a row vector, can be similarly transformed into the $K$-matrix dipole vector by right-multiplying it by the matrix ${\cal C}^{-1}$, i.e. $d^{(K,b)}_{i''}= \sum_i d^{(b)}_{i'} ({\cal C}^{-1})_{i',i''}$. The reaction matrix $K$ formed at this point will typically be almost symmetric but possibly not quite, and numerical stability is improved slightly if this $N \times N$ matrix $K^{(b)}$ is now symmetrized manually, i.e. by averaging $K^{(b)}$ and its transpose ${\tilde K}^{(b)}$:   $K^{(b)} \rightarrow \frac{1}{2}(K^{(b)}+{\tilde K}^{(b)})$.

\subsubsection{Enforcing exponential decay in strongly-closed ionization channels}
For each set $b$ of $N$ independent solutions formed through the above procedure, in the region $R<R_0$, the normalized vibrational wavefunctions fall into two qualitatively different groups.  In group $P$ are all the low vibrational eigenstates that have negligible amplitude or derivative on the dissociative surface $R_0$, group $Q$ consists of the set of relatively high $v^+$ that have appreciable amplitude or derivative at $R_0$.  The number $N_P$ of channels in the $P$ subspace is typically around 8 for each $N^+ = 0,2$ for $J=1$, giving $N_P \approx 16$ and $N_Q = N - N_P$.  The enforcement of exponential decay in the electronic wavefunctions of the $Q$ subspace is carried out next, through the standard equations of MQDT, which are Eqs. (17)-(23) of Ref.\cite{hvizdovs2025competing}.  The following matrix notation is convenient, in which the electronic radial wavefunctions in each independent solution $i'$ at $r>r_0$ are arranged to form one column of the solution matrix $M(r)$.  That is, $\Psi_{i'} = \sum \Phi_i M_{ii'}(r)$, and the partitioned form of the radial solution matrix looks like 
\begin{equation}
  \label{eq-partition}
      M(r) = 
    \begin{bmatrix}
        M_{PP}=f_{P}-g_{P}K_{PP} & M_{PQ}=-g_{P}K_{PQ} \\
        M_{QP}=-g_{Q}K_{QP} &M_{QQ}=f_{Q}-g_{Q}K_{QQ} \\
    \end{bmatrix}.
\end{equation}
Taking linear combinations of these states, intended in particular to enforce exponential decay and produce a smaller set of $N_P$ solutions for each $b$, is represented mathematically by right-multiplying Eq.(14) by an $r$-independent matrix given by:
\begin{equation}
\label{eq-channel_elimination_A}
    \left( 
    \begin{array}{c}
    {\bf 1}_{PP} \\ 
    A^{(b)}_{QP}%
    \end{array}%
    \right) =\left( 
    \begin{array}{c}
    {\bf 1}_{PP} \\ 
    -(K^{(b)}_{QQ}+\tan \beta_{Q})^{-1}K^{(b)}_{QP}
    \end{array}%
    \right) 
\end{equation}
giving for our new solution matrix the following:
\begin{align}
    \nonumber
    {\mathcal F}^{(b)}(r) &= \left( 
   \begin{array}{cc}
    f_{P}-g_{P}K^{(b)}_{PP} & -g_{P}K^{(b)}_{PQ} \\ 
    -g_{Q}K_{QP} & f_{Q}-g_{Q}K^{(b)}_{QQ}%
    \end{array}%
    \right) \left( 
    \begin{array}{c}
    {\bf 1}_{PP} \\ 
    A^{(b)}_{QP}%
    \end{array}%
    \right) \\
    \label{eq-channel_elimination_F}
    &= \left( 
    \begin{array}{c}
    f_{P}(r)-g_{P}(r)\mathcal{K}^{(b)}_{PP} \\ 
    W^{(b)}_{Q}(r)Z^{(b)}_{QP}%
    \end{array}%
    \right). 
\end{align}
The first MQDT ``pre-elimination'' of strongly closed channels, described by Eq.~(\ref{eq-channel_elimination_F}), involves a reduction of the channel space to a reaction matrix ${\mathcal{K}_{PP}}$ involving a smaller set $\{P\}$ of ``open or weakly-closed channels''.  The familiar equation is
\begin{equation}
\label{eq-channel_elimination_K_basic2}
    {\mathcal{K}^{(b)}_{PP}}= K^{(b)}_{PP}-K^{(b)}_{PQ}(K^{(b)}_{QQ}+\tan{\beta_Q})^{-1} K^{(b)}_{QP},
\end{equation}
where $\tan{\beta_Q}=\tan{\pi(\nu_Q-l)}$ is a diagonal matrix needed to eliminate the exponentially growing terms. Because the ionization thresholds $i\equiv \{v^+,N^+\}$ in the $Q$ subspace depend on the boundary condition $b$ imposed on the vibrational channel functions, $\nu_Q$ should be understood to have a $b$ superscript implied, but for brevity it is not written here explicitly. While this step eliminates the exponentially growing terms in the $Q$-channel space, the coefficients of the exponentially decaying ``energy normalized'' Whittaker functions of $r$ in the $Q$-channels are needed for some of the steps below. 
Here $W^{(b)}_Q(r)$ is an ``energy-normalized'' Whittaker Coulomb solution, normalized as in Eq.~(2.45) of Ref.\cite{Aymar_Greene_LKoenig_1996}.  The amplitude multiplying the Whittaker function is given by the following algebra:
\begin{align}
\label{eq-Zmat}
    &Z^{(b)}_{QP}= \sin{\beta_Q} K^{(b)}_{QP} 
     +(\cos{\beta_Q}-\sin{\beta_Q} K^{(b)}_{QQ}) (K^{(b)}_{QQ}+\tan{\beta_Q})^{-1}K^{(b)}_{QP}.
\end{align}
After this step is completed, each of the two sets of $N_P$ independent solutions can be written in the following form, essentially an expanded way of writing out Eq.~(\ref{eq-channel_elimination_F}) in detail, for $i'\in P$:
  \begin{align}
  \label{eq-preelim}
     \Psi^{(b)}_{i'}(E,r,R) = &\sum_{i\in P} 
    \Phi_i
    \bigg(f(\epsilon_{i},\ell_{i},r) \delta_{i i'} 
    - g(\epsilon_{i},\ell_{i},r) \mathcal{K}^{(b)}_{i i'}(E) \bigg)  \nonumber \\
   &+ \sum_{i\in Q}\Phi_i^{(b)} W_i^{(b)}(r) Z_{i,i'}^{(b)},
  \end{align}  
noting that $\Phi_i$ are assumed to be independent of the boundary parameter $b$ in the $P$ space .

The same strongly-closed channel elimination procedure must of course be applied to the transition dipole matrix element row vectors $d^{(b)}$.  This step creates two energy-dependent row vectors, each of length $N_P$ by right-multiplying with the transformation matrix in Eq.~(\ref{eq-channel_elimination_A}):
\begin{equation}
    {\cal D}^{(b)}_P=d^{(b)}_P-d^{(b)}_Q (K_{QQ}+\tan{\beta_Q})^{-1} K_{QP}
\end{equation}

It should be kept in mind that at this point there are still weakly closed channels in set $P$, whose exponential divergence at $r\rightarrow \infty$ will be eliminated later.  But aside from this divergence, the solutions created up to now will be solutions of the model Schr\"odinger equation throughout the region defined by $r$ right outside the interaction range and $R<R_0$.  However, to determine the portions of the full scattering matrix, we must extract the information at the dissociative boundary $R=R_0$ in each solution obtained up to now.  That information is contained entirely and exclusively in the second term of Eq.~(\ref{eq-preelim}), because all of the vibrational wavefunctions in $\Phi_P$ have negligible value and derivative at $R_0$.

The remaining full details concerning that extraction of the dissociative portions of the wavefunctions and reaction matrix, and the combination of the overcomplete set of solutions into a single unified set of independent solutions characterized by a single MQDT reaction matrix, are laid out in the Appendix.  These steps closely follow the logic of our treatment that was presented in Ref.\cite{hvizdovs2025competing}, where those steps are explained in greater detail.

Once the steps described in the Appendix have been carried out, the calculation has produced a reaction matrix in the space of open and weakly-closed ionization channels $v^+,N^+$ and with energetically open dissociation channels $d$.  This matrix is relatively smooth {\it because} the final boundary condition in the weakly closed channels has not yet been imposed.  That is now carried out with the small reaction matrix $\mathcal K$ to obtain the highly energy-dependent final physical reaction matrix ${\mathcal K}^{\rm phys}$, for which all exponential growth at $r\rightarrow \infty$ will have been imposed.  This involves right-multiplication by one final ``closed-channel elimination'' matrix as in Eq.~(\ref{eq-channel_elimination_A}).  These remaining steps are now a straightforward implementation of standard MQDT formulas, as formulated in Sec.II of \cite{Aymar_Greene_LKoenig_1996}.  In brief, consider a calculation at one specific energy $E$ lying in the vicinity of the lower few ionization thresholds, and let $p$ denote the set of energetically open ionization and dissociation channels, while $q$ represents the set of closed channels.  (For the energy range considered in the present study, all of the closed channels considered will be ionization channels, but if higher energies are considered at some point, one can broaden the scope of this discussion and include weakly closed dissociation channels in the set $q$.)

\begin{equation}
    {\cal K}^{\rm phys}={\cal K}_{pp} - {\cal K}_{pq}({\cal K}_{qq}+\tan{\beta_q})^{-1} {\cal K}_{qp}
\end{equation}
and
\begin{equation}
    \label{eq-D_final_elim}
    {\cal D}^{K^{\rm phys}} = {\cal D}_p-{\cal D}_q ({\cal K}_{qq}+\tan{\beta_q})^{-1} {\cal K}_{qp}
\end{equation}
Then in order to compute observables such as the dissociative recombination cross section, photoionization and photodissociation cross sections, it is helpful to carry out one final transformation to obtain the Hermitian conjugate of the physical scattering matrix ${\cal S}^\dagger$ and the corresponding energy eigenstates that asymptotically obey the incoming-wave boundary condition.  That transformation right-multiplies the preceding solution by $(1+i {\cal K}^{\rm phys})^{-1}$, giving the matrix
\begin{equation}
    {\cal S}^{\dagger {\rm phys}}=\frac{1-i{\cal K}^{\rm phys}}{1+i{\cal K}^{\rm phys}}
\end{equation}
and the corresponding physical dipole amplitudes are
\begin{equation}
    {\cal D}^{{\cal S}^{\dagger {\rm phys}}} =  {\cal D}^{K^{\rm phys}}(1+i {\cal K}^{\rm phys})^{-1}
\end{equation}
The partial photofragmentation cross section into channel $i$, regardless of whether it is a photoionization or a photodissociation channel, is now given in atomic units by:
\begin{equation}
    \label{eq-photo-cross-section}
    \sigma_i = \frac{4\pi^2 \omega \alpha}{3(2J_0+1)} |{\cal D}^{{\cal S}^{\dagger {\rm phys}}}_i|^2,
\end{equation}
where $\alpha \approx 1/137.036$ is the fine-structure constant, and $\omega$ is the photon frequency.  Similarly, the partial inelastic or reactive scattering cross section, e.g. for processes such as dissociative recombination, vibrational excitation, associative ionization, etc., starting from incident channel $i'$ and ending in a different channel $i$, is equal to:
\begin{equation}
    \label{eq-scatt-cross-section}
    \sigma_{i,i'} = \frac{\pi}{2 \epsilon_{i'}(2{N^+}'+1)} \sum_J (2J+1) |{\cal S}^{{\rm phys}}_{ii'}|^2.
\end{equation}
This formula considers only the contribution of a single partial wave $\ell,J$, consistent with the model treated throught the present study, but the other $\ell,J$ channels can be incorporated when appropriate.

\section{Results}

\begin{figure}[tbh]
    \centering
    \includegraphics[width=0.8\linewidth]{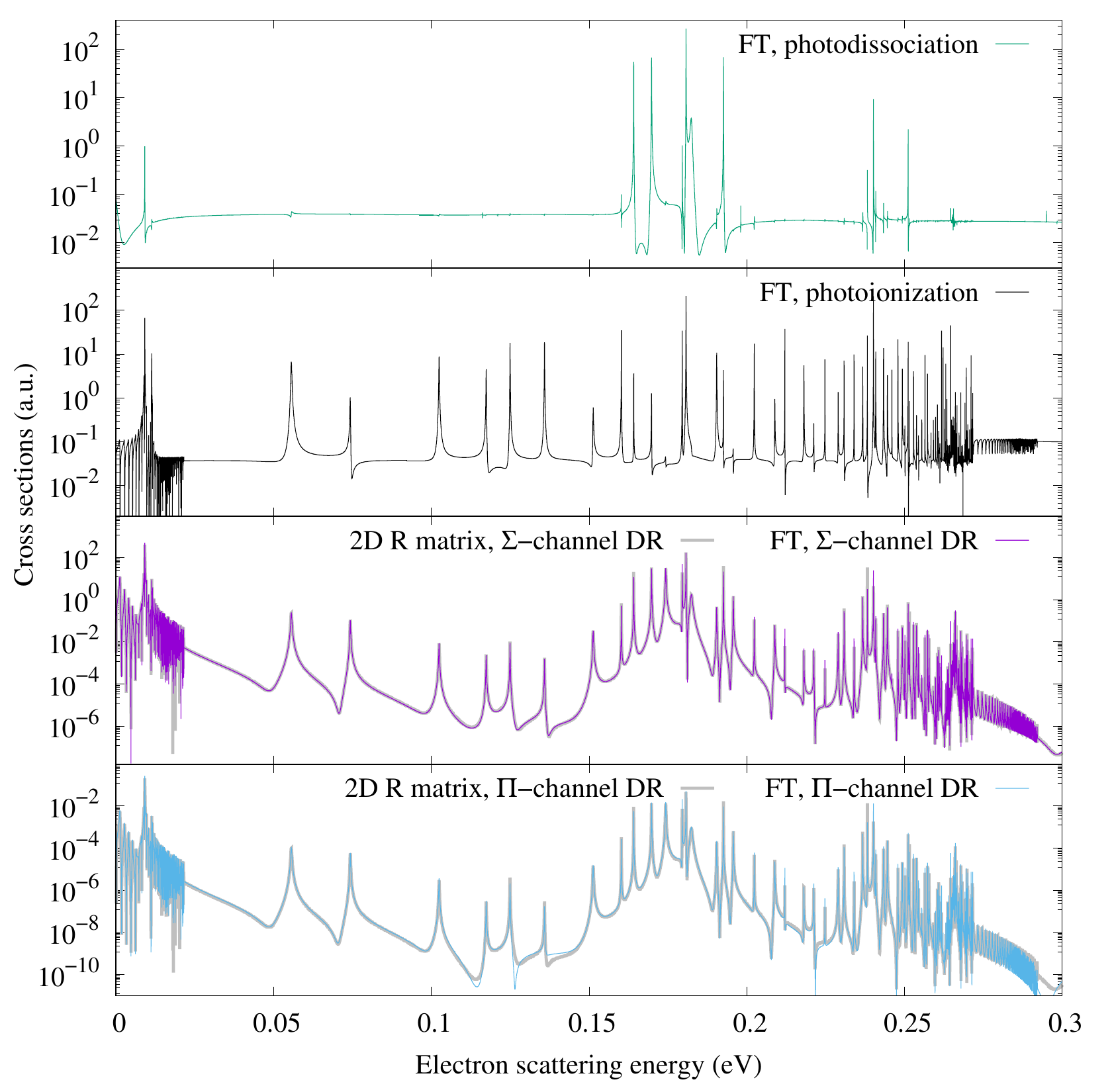}
    \caption{Cross section comparison between DR $^1\Pi_u$ output channel (green curve), DR $^1\Sigma_u^+$ output channel (purple curve), photoionization (black curve) and photodissociation (green curve). The 2D R matrix benchmark for DR is present as the background thick grey curve.
    }
    \label{fig-DR_benchmark1}
\end{figure}

Fig.~\ref{fig-DR_benchmark1} show some of our main results from this study of the final $J=1$, odd-parity state of para-H$_2$, for final states $^1\Sigma_u^+$ and $^1\Pi_u$, respectively.  On the figures a rich resonance pattern is observed across the energy range from 0 to 0.3 eV above the lowest ionization threshold, and for each resonance different lineshapes occur for the three different observables displayed: total photoionization (PI) of the H$_2$ rovibrational ground state, total photodissociation (PD), and dissociative recombination (DR) of electrons incident on the $v^+=0,N^+=0$ ionic ground state.  First of all, dissociative recombination has the smallest cross section over most of this energy range, and dissociation into the B$'$ $^1\Sigma_u^+$ state of H$_2$ (upper figure) dominates over the partial cross section into the C $^1\Pi_u^+$ state by around 3 orders of magnitude.  A second major takeaway is that our approximate EDFT treatment of the DR process is in quite good agreement with our essentially exact 2D R-matrix calculation, throughout this energy range; the main differences become apparent predominantly in regions where the cross section is especially small, i.e. $\lesssim 10^{-6}$ a.u.  

At first glance, when viewed on a logarithmic scale as in Fig.~\ref{fig-DR_benchmark1}, the resonances in photoionization show strong correlation with the DR resonances; the DR resonances in this system are seen to be usually better correlated with photoionization than with the photodissociation resonances. An example of this is the region from 0.03 eV to 0.15 eV, where this symmetry shows 7 strong photoionization resonances, each with a clear signature in the DR spectrum as well.  In this energy range, no photodissociation resonances are even visible in Fig.~\ref{fig-DR_benchmark1}.  This poorer correspondence between DR resonances and photodissociation resonances, in comparison with photoionization resonances, was initially a surprise, since photoionization only requires an electron to be excited, while all dissociative processes require a transmission of that energy to the nuclear motion.  On the other hand, note that the expected stronger correlation between DR and photodissociation is in fact observed in the regions of complex resonances, where a low $n$, high $v^+$ perturbing level is immersed in a high $n$, low $v^+$ Rydberg series, as in the regions around 0.01 eV, 0.16-0.20 eV,  and 0.24 eV.  These are in fact the only regions below 0.3 eV where the total DR cross section exceeds 1 a.u., and in each of those regions a strong correlation with strong photodissociation resonances is apparent.  The dominant role of complex resonances as a leading conduit for DR in molecules dominated by the indirect DR mechanism was stressed in a previous study of LiH$^+$ dissociative recombination in Refs.\cite{Curik_Greene_PRL_2007,Curik_Greene_MP_2007}  The reason for this dominance of complex resonances for DR is physically intuitive:  the low $v^+$ parts of the wavefunctions are relatively easily excited by a photon or an electron collision, and their strong coupling to a perturbing level in a high vibrational state, which is the characteristic feature of a complex resonance, efficiently transmits that excitation to the nuclear motion.

\begin{figure}[tbh]
    \centering
    \includegraphics[width=0.7\linewidth]{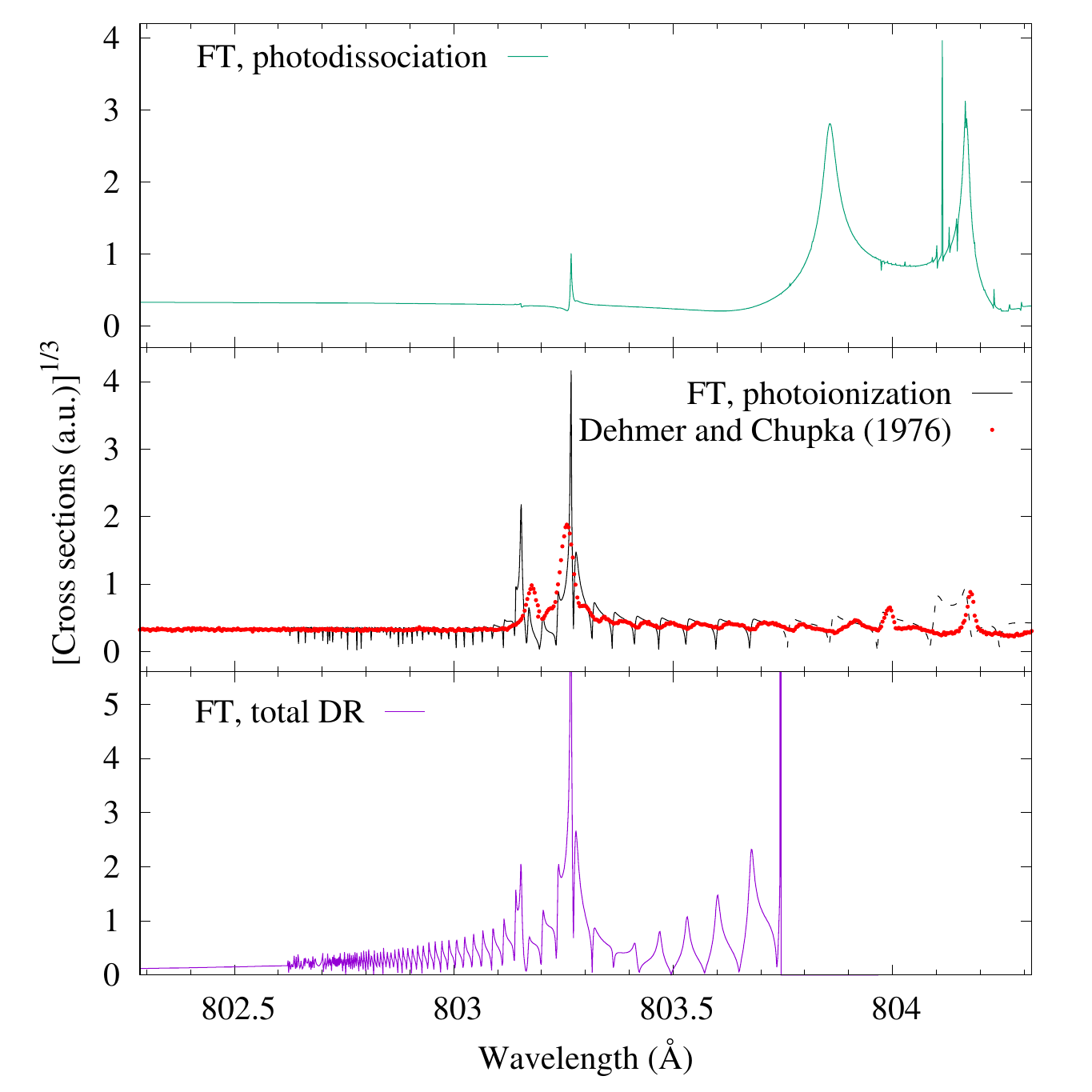}
    \caption{Cross section cube root comparison close to ground-vibrational-state threshold (which is at approximately 803.746 \AA). The cross sections shown are total DR (purple curve), photoionization (black curve), photodissociation (green curve) and experimental photoionization data of Dehmer and Chupka (red data points).  As Ref.\cite{Dehmer_Chupka} points out, their photoionization spectrum observes signal even at energies slightly below the lowest ionization threshold, and they attribute this to the likely presence of weak electric fields.  The dashed curve in the PI middle panel of the figure shows our mock photoionization cross section obtained by artificially opening that lowest ionization channel, which simulates the signal expected if very high photon-excited Rydberg states are field ionized in the experiment.  In the DR spectrum, an infinite spike marks the lowest ionization threshold, an expected divergence because of the $1/k^2$ factor in the DR cross section formula.  The two broad PD resonances below threshold were previously observed by Ref.\cite{HerzbergJungen}, and their classifications are discussed in the text.
    }
    \label{fig-photionization_low}
\end{figure}
The energy range expanded in Fig.~\ref{fig-photionization_low}, close to the $v^+=0,N^+=0$ ionization threshold, has been a remarkable and iconic energy range, since the experimental photoabsorption measurements of Herzberg, of photoionization by Dehmer and Chupka\cite{Dehmer_Chupka}, and reproduced accurately in theory for the first time and apparently the only time until now by Jungen and Dill\cite{Jungen_Dill_JCP_1980}. See especially Figs.~3, 5, and 6 of their 1980 masterpiece. The depiction of this double complex resonance in Fig.~\ref{fig-photionization_low} shows a number of interesting features associated with the two lower-$n$, perturbers ($7p\pi,v=1$ and $5p\pi,v=2$) that occur between 803.1 and 803.3 \AA{}.  One sees again the key role of the complex resonances as the dominant regions where dissociative recombination is strong, one of which is centered near 803.27 \AA{} in good agreement with the tallest experimental peak and the other of which lies at 803.14 \AA{} threshold and sits at a higher energy than the experimental peak.  The photoionization spectrum shows two main peaks near 803.1 \AA{} and 803.26 \AA{}, which appears to be two complex resonances interfering with each other, and with the many rotationally-autoionizing Rydberg levels that are converging to the $v^+=0,N^+=2$ first excited ionization threshold. Both complex resonances show parallel enhancements in the DR cross section in that energy range, though the one at lower energy is far stronger.  Their interplay with the rotationally-autoionizing states produces a stegosaurus-like structure in the DR spectrum around 803.5 \AA{}.  That strongest and narrowest resonance also leads to resonant enhancement of the total PD spectrum, and two broad PD resonances are also seen that peak below the lowest ionization threshold.  We find it encouraging that our crude model with rms errors of order 0.01 in the body frame quantum defect functions is able, when combined with the EDFT method, to reproduce quite well the major features in the experimental photoionization spectrum, and predict new features in DR and PD that have apparently not previously been computed nor measured experimentally.

One curious aspect of the experimental photoionization spectrum is that ionization is observed even below the ionization threshold.  Dehmer and Chupka in Ref.\cite{Dehmer_Chupka} speculate that the very high Rydberg states just below the ionization threshold might get ionized by stray electric fields present in their apparatus.  But our calculations do predict two very strong PD resonances and this is an energy range where predissociated photoabsorption resonances were observed by Herzberg and Jungen in Ref.\cite{HerzbergJungen}, classified as $3p\pi,v=6$ at 803.872 \AA{} or 124398 cm$^{-1}$, and $4p\sigma, v=4$ at 804.128 \AA{} or 124358 cm$^{-1}$.  Our two computed strong and broad below-threshold PD peaks in Fig.\ref{fig-photionization_low} appear to correspond to those two resonances, and our positions are shifted from the Ref.\cite{HerzbergJungen} photographic plate positions by 2 cm$^{-1}$ and -6 cm$^{-1}$, respectively.  The position of the computed PD resonance at $\sim$804.17 \AA{} lines up approximately with the lower energy below-threshold ``photoionization'' resonance in the Ref.\cite{Dehmer_Chupka} experimental spectrum plotted in Fig.\ref{fig-photionization_low}.  The dense, high-$n$ members of the Rydberg series converging to the first excited threshold $v^+=0,N^+=2$ are not visible at all in the PD spectrum.   `

\begin{figure}[tbh]
    \centering
    \includegraphics[width=0.7\linewidth]{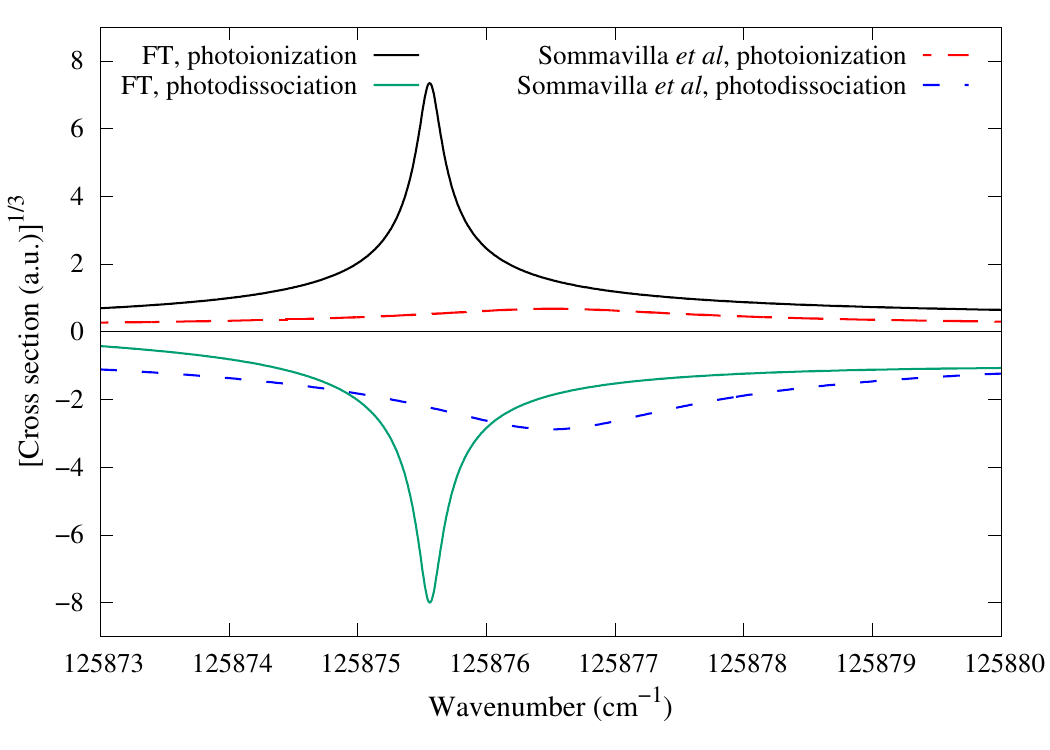}
    \caption{Photoionization (black curve) and photodissociation (multiplied by -1, green curve) cross sections near 125875 cm$^{-1}$ produced by our model compared with the PI and PD cross sections calculated in Ref.\cite{Sommavilla_Merkt_Zsolt_Jungen_2016jcp} and shown with dashed curves. The cross sections are all showed on a cube root scale in order to assess the differences and similarities more clearly. The experimental spectrum from Ref.\cite{Sommavilla_Merkt_Zsolt_Jungen_2016jcp} in this region is not shown here because it is mixed with an obscuring R(1) transition at nearly the same energy.  While the R(0) resonance position from our model EDFT calculation is within 1 cm$^{-1}$ of the Ref.\cite{Sommavilla_Merkt_Zsolt_Jungen_2016jcp} computed position, the resonance width and branching ratio disagree strongly with that 2016 study, presumably indicating a limitation of our model.
    }
    \label{fig-Somavilla}
\end{figure}
A particularly sensitive test of one highly predissociated photoabsorption resonance is shown in Fig.~\ref{fig-Somavilla}.  Here our EDFT calculation based on the body-frame quantum defects from our simplified model is seen to predict the position of the resonance with an error of the order of 1 cm$^{-1}$.  This energy range was studied in the 2016 study by Sommavilla {\it et al.} \cite{Sommavilla_Merkt_Zsolt_Jungen_2016jcp} and in the experiment this region was shown to consist of an R(0) resonance that strongly overlaps an R(1) resonance.  Because our calculation is only for the final $J=1$ state reached from the para-H$_2$ ground state, we compare only with the R(0) spectrum in this region, for both the PI and PD observables, determined in the highly accurate calculations carried out by the theorists in that study. 

\begin{table}[tbh]
\caption{\label{tab-somma}Comparison between the $R(0)$ photoionization transition wavenumbers $\nu$ taken from \citet{Sommavilla_Merkt_Zsolt_Jungen_2016jcp} and positions of presently calculated closed-channel resonances in the DR cross sections using our model Hamiltonian.  Differences listed in the last column are only displayed to 0.1 cm$^{-1}$ precision.  As part of our benchmarking, our resonance positions are computed with both the essentially exact 2D R-matrix solution of our model (final column) and with the EDFT approximation.  While the latter are not shown, we find that the EDFT resonance positions agree with the R-matrix calculated positions to within 0.1 cm$^{-1}$. \\ 
$^*$For the two unobserved transitions (n.o.), the calculated values \cite{Sommavilla_Merkt_Zsolt_Jungen_2016jcp} are listed in the experimental column.}
\begin{center}
\begin{tabular}{rrrr}
\hline
 & \multicolumn{2}{c}{\citet{Sommavilla_Merkt_Zsolt_Jungen_2016jcp}} & \\
 \cline{2-3}
Assignment & Exp. & Calc.$-$Exp. & present$-$Exp.$^*$\cite{Sommavilla_Merkt_Zsolt_Jungen_2016jcp}  \\
\hline
$10p2, v=1$ & 125638.50 & 0.06 & -0.4  \\
$11p0, v=1$ &    710.42 & 0.06 & -0.3  \\
$5p\sigma, v=3$ & 730.04& 0.18 & 11.6  \\
$6p\pi, v=2$ & 781.31 & n.o. & 6.0  \\
$12p0, v=1$  & 823.46 & 0.03 & 0.0  \\
$11p1, v=1$  & 865.78 & -0.06 & 0.0  \\
$3p\pi, v=7$ & 876.28 & -0.27 & -0.9  \\
$4p\sigma, v=5$ & 889.70 & -1.1\phantom{0} & -0.4  \\
$13p0, v=1$ & 953.91 & 0.09 & 0.0  \\
$4p\pi, v=4$ & 970.23 & 0.14 & 1.0  \\
$12p2, v=1$ & 996.52  & n.o. & -0.1 \\
\hline
\end{tabular}
\end{center}
\end{table}

It is apparent that our model produces a resonance for this state, classified as $3p\pi,v=7$, which is significantly narrower than the accurate MQDT calculation of Ref.\cite{Sommavilla_Merkt_Zsolt_Jungen_2016jcp}. Moreover, our calculation predicts nearly equal PD and PI cross sections at the resonance maximum, whereas the MQDT calculation of their published article for the R(0) symmetry shows strong dominance of the photodissociation channel. This is one of the regions showing the largest discrepancy of our results from previous experiment or accurate theory, over this entire energy range up to $\sim$ 1 eV above the ionization threshold.  It suggests that for some unusually sensitive regions in the spectrum, the errors of order $\Delta \mu_\Sigma \sim 0.01 $ in the quantum defect functions predicted by our model are too large to correctly describe this resonance decay properties, which might signal strong interference between competing channels.  We suspect that this could be remedied by improving our 2D model Hamiltonian and ensuring that the known, highly accurate Born-Oppenheimer potential curves are reproduced with higher precision, but this hypothesis has not yet been tested.

We further explored the energy window surrounding the $3p\pi,v=7$ state in which 10 more $R(0)$ photoionization transitions were reported by \citet{Sommavilla_Merkt_Zsolt_Jungen_2016jcp}. These transitions are compared with positions of closed-channel resonances in the presently computed DR cross sections. The comparison in Tab.~\ref{tab-somma} shows a good agreement for the high $n$ (low $v^+$) states indicating that our near-zero-energy quantum defects are reasonably accurate. However, their negative-energy dependence leads to larger discrepancies for low $n$ (high $v^+$) states.  We also used the calculated resonance positions as a test of the accuracy of the energy-dependent frame transformation, and find that they agree with the 2D R-matrix positions at the level of 0.1 cm$^{-1}$.
\begin{figure}[bth]
    \centering
    \includegraphics[width=0.7\linewidth]{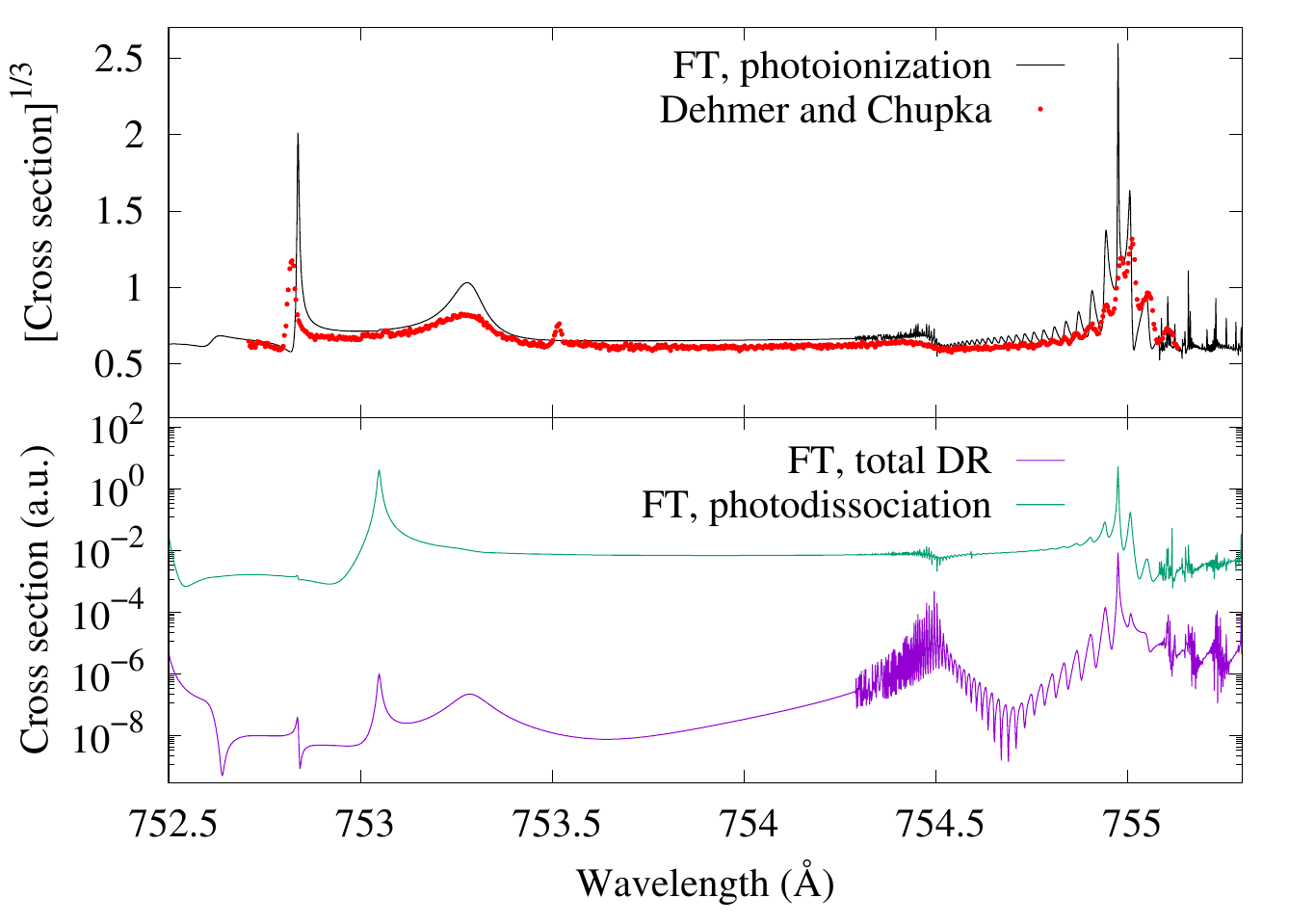}
    \caption{Cross section comparison of total DR (purple curve), photodissociation (green curve), photoionization (black curve) and experimental data (red points).  The PD and PI cross sections are displayed on a cube root scale.  Note that we have followed the recommendation of Ref.\cite{Jungen_Raoult_1981}, which carried out the first successful calculation of this photoionization spectrum, and have similarly shifted the experimental data in this energy range by -0.068 \AA{}, which Jungen and Raoult attribute to a stepping motor slippage error that occurred during the data acquisition of Ref.\cite{Dehmer_Chupka}.
    }
    \label{fig-photionization_high}
\end{figure}

A final example spectrum is shown in Fig.~\ref{fig-photionization_high}. Here there is again an impressive data set from a half-century ago, published by Dehmer and Chupka\cite{Dehmer_Chupka}, and also a fine, early MQDT calculation based on energy-independent quantum defect functions, that was carried out by Jungen and Raoult\cite{Jungen_Raoult_1981}.  Their MQDT calculation is not shown here, but the reader is encouraged to take a look at Fig.~4 of Ref.\cite{Jungen_Raoult_1981}, where they have very accurately reproduced the photoionization spectrum in the energy range of Fig.~\ref{fig-photionization_high}, including the complex resonance near 755 \AA{} and other more subtle features in the experimental spectrum as well.  This figure demonstrates that our EDFT calculation based on the quantum defect functions from our model Hamiltonian also accurately describes the PI spectrum.  

In addition to the PI spectrum, our calculation predicts the photodissociation spectrum, with resonant PD enhancements at the complex resonance and also at a large resonance near 753.05 \AA{} that is almost invisible in the PI spectrum.  If one looks very closely at the PI spectrum there, a tiny inflection is in fact barely visible in our computed PI spectrum at that energy.  While the photoionization is too weak to show up in the PI spectrum computed in Ref.\cite{Jungen_Raoult_1981}, that resonance has been observed in photoabsorption by Glass-Maujean and Schmoranzer\cite{glass2022absorption} at a photon wavenumber of 132792.1 cm$^{-1}$ relative to the ground state with a decay width quoted at 3 cm$^{-1}$. The position of that strong photodissociation resonance, classified in Ref.\cite{glass2022absorption} as $3 p \pi, v=14$, is computed to lie 1.2 cm$^{-1}$ higher in both our EDFT and 2D R-matrix calculations.  In the DR spectrum, recombination is predicted to be enhanced at the complex resonance centered at 755 \AA{}, and also at another complex resonance centered at 754.5 \AA{} that has only a tiny signature in the PI spectrum.  Note that the experimental resonance at 753.5 \AA{} was assigned in Ref.\cite{Jungen_Raoult_1981} to a different initial state $J''=1$, and so it is not present in our computed spectrum.

\section{Conclusions}
The present study has shown that the energy-dependent frame transformation theory, combined with a Jungen-Ross-style treatment of dissociation along the lines of our reformulation presented in Ref.\cite{hvizdovs2025competing}, is capable of describing the complicated multichannel rovibrational interactions in molecular hydrogen. While the DR results shown in Fig.\ref{fig-DR_benchmark1} have been computed using both the 2D R-matrix method and the EDFT treatment, the PI and PD cross sections have only been computed using the EDFT approach. The EDFT has been benchmarked carefully against an essentially exact solution for our model Hamiltonian, showing that the two methods agree quantitatively, except for minor differences in regions where the DR cross sections are extremely small.  The model itself, when implemented with the approximate EDFT approach, describes most of the spectrum in terms of resonance positions and strengths to the level of 1 cm$^{-1}$, with only rare exceptions.  The present study also adds further confirmation to the finding in Ref.\cite{Curik_Greene_PRL_2007} that complex Rydberg resonances provide an especially efficient DR mechanism, at least for systems where the indirect DR process dominates.

Quite apart from the primary focus of this article, which has been to test this theoretical framework quantitatively for multiple observables in the near-threshold energy range of H$_2$, the study also shows the power of analyzing collisional observables in a spectral region dominated by resonances, with an eye aiming towards the extreme precision that is routinely achieved in photon absorption experiments.  Tremendous progress has been made in recent decades in such spectroscopic observations, e.g. by Merkt, Ubachs, Beyer, Glass-Maujean, as well as other groups, and it is our hope that this powerful progress can also be harnessed to better understand electron collision experiments going forward.

\section{Acknowledgments}
We  thank Christian Jungen for discussions and advice, and for providing tabulated data from Ref.\cite{Sommavilla_Merkt_Zsolt_Jungen_2016jcp}. Access provided by Patricia M. Dehmer to the original tabulated data of Ref.\cite{Dehmer_Chupka} is appreciated. Conversations with Ioan Schneider and Zsolt Mezei have also been helpful.  The Purdue portion of this work was supported by the Department of Energy, Office of Science, Office of Basic Energy Sciences, Award No. SC0010545. R.\v{C}. acknowledges support of the Czech Science Foundation (Grant No.\ GACR 21-12598S).
D.H. acknoledges the support of Universit\'{e} Le Havre Normandie via the French Programme PIA4 and la R\'{e}gion Normandie.

\section*{Author declarations}
The authors have no relevant competing interests to report.

\section*{Data Availability Statement}
The authors confirm that the data supporting the findings of this study are available within the article and raw digital data are available from the corresponding author, D.H., upon reasonable request.

\bibliographystyle{**unsrturl**}
\bibliography{bibliography}

\appendix

\section{Construction of the dissociative parts of the wavefunctions}

As was stressed by Fano in Ref.\cite{Fano_PRA_1970}, Hund's case (d) is relevant when an electron moves far away from the ion and the ionic angular momentum $N^+$ becomes an appropriate channel quanum number.  When the molecule begins to dissociate at an energy within 1 eV of the ionic ground state energy, however, the outermost electron is tightly bound to a nucleus and Hund's case (b) becomes more appropriate in the dissociative region $R\ge R_0$ of the position space.  Accordingly, we start by transforming the second term of Eq.~(\ref{eq-preelim}) from Hund's case (d) to Hund's case (b) where $\Lambda$ is a good quantum number in addition to the principal quantum number $n$.
\begin{align}
    \nonumber
    &\sum_{i\in Q}\Phi_i^{(b)} W_i^{(b)}(r) Z_{i,i'}^{(b)} \\
    & =  \sum_{\{v^+ N^+\}\in Q} |(\ell N^+)J\rangle \chi_{v^+N^+}^{(b)}(R) W_{v^+N^+}^{(b)}(r) Z_{{v^+N^+,i'}}^{{(b)}} \label{eq-strng-recoup} \\
    & \approx \sum_{n\Lambda} |\ell\Lambda J \rangle \phi^{\rm BO}_{n\Lambda}(r;R)  P^{(b)}_{n\Lambda,i'}(R),\ {\rm \ for\ }R \sim R_0.  \label{eq-BO}
\end{align}
Our assumption, analogous to that made in Refs.\cite{Jungen_Ross_1997,hvizdovs2025competing}, is that the $i'-{\rm th}$ independent solution in Eq.~(\ref{eq-strng-recoup}) should be approximately equal at $R=R_0$ to a combination of electronic Born-Oppenheimer solutions $\phi^{\rm BO}_{n\Lambda}$ at $R_0$, to within a normalization factor.  In the energy range considered here, for our greatly simplified model of the {\it ungerade} states of H$_2$, two $^1\Sigma_u$ dissociating potential curves are energetically open in this energy range, namely the $n=1$ and $n=2$ dissociation channels, but only one $^1\Pi_u$ channel is open, which dissociates to $H(n=2)$.  In the real-world H$_2$ molecule, of course, there is no singlet ungerade state that dissociates to the $n=1$ threshold.  The lowest state of that symmetry is in fact the $B$-state, which in reality dissociates to $n=2$.  We view our lowest ungerade state loosely as a very poor representation of that $B$-state of H$_2$, and consequently any results obtained that significantly involve that lowest ungerade state in our model should not be taken seriously.  For the physics near the ionization threshold that interests us, this unphysical representation of the actual molecular $B$-state should not matter.

 Now the idea is to take seriously the equality between Eqs.~(\ref{eq-BO}) and (\ref{eq-strng-recoup}), and also the equality of the radial $R$ derivatives of those two equations.  Then the key quantities $P^{(b)}_{n\Lambda,i'}(R)$ can be determined for the three ungerade dissociation channels relevant in this targeted energy range, e.g. by  projecting Eq.~(\ref{eq-strng-recoup}) onto the orthonormal Born-Oppenheimer eigenfunction 
$|n\ell \Lambda J \rangle \Phi^{\rm BO}_{n\Lambda}(r;R)$, or by least-squares fitting of the $r$-dependence in Eq.~(\ref{eq-BO}) to the expected linear combination of normalized Whittaker solutions with the Born-Oppenheimer bound state effective quantum numbers at the matching radius $R_0$, $\nu_{n\Lambda}(R_0)$.

After that analysis just described has been carried out, the quantities extracted from this first step of elimination of strongly-closed ionization channels will be $P^{(b)}_{n\Lambda,i'}(R_0)$ and its $R$-derivative, $P^{(b)'}_{n\Lambda,i'}(R)$ for each of the remaining $N_P$ independent solutions $i'$.  Then, after those quantities have been obtained, they can be matched to a linear combination of two energy-normalized solutions in the open dissociation channels.  This matching involves the following equation and its $R$-derivative:
\begin{equation}
\label{eq-dissociative_rows}
     P^{(b)}_{di'}(E,R_0) = {\bar F}_{E_d}(R_0) I^{(b)}_{d i'}-{\bar G}_{E_d}(R_0) J^{(b)}_{di'},
\end{equation}
The next step is to solve for the constant matrices that will characterize the $i'$-th independent solution for boundary parameter $b$ at distances $R\ge R_0$, using:
\begin{equation}
\label{eq-dissociation_IJ}
\begin{aligned}
    I^{(b)}_{d i'} = \frac{\mathcal{W}[P^{(b)}_{di'},{\bar G}_{E_d}]}{\mathcal{W}[{\bar F}_{E_d},{\bar G}_{E_d}]}, \\
    J^{(b)}_{d i'} = \frac{\mathcal{W}[P^{(b)}_{di'},{\bar F}_{E_d}]}{\mathcal{W}[{\bar F}_{E_d},{\bar G}_{E_d}]},
\end{aligned}
\end{equation}
where $\mathcal{W}[\cdot,\cdot]$ is a Wronskian in the $R$-coordinate evaluated at $R_0$.
Once the coefficients of those linear combinations are known, we have in effect added three extra rows to our solution matrix (two $\Sigma$ dissociation channels and one $\Pi$ dissociation channel), enlarging the solution matrix for each $b$ to the dimension $(N_P+3) \times N_P$.  Each $b$ now has its corresponding set of $i'=1$ to $N_P$ solutions, with the following channel structure outside the reaction volume:
\begin{align}
    \nonumber
      \Psi^{(b)}_{i'}= &\sum_{i \in P} \Phi_i  {\bigg (} f_i(r) \delta_{i{i'}} -g_i(r) {\mathcal K}^{(b)}_{i{i'}} {\bigg )}, \ \ r>r_1 \ {\rm and}\ R<R_0\\
    \label{eq-wavefunction_joined_full}
    &+ \sum_d \ket{\phi_{d}^{\text{BO}}(r;R_0)} {\bigg (} {\bar F}_{E_d}(R) I^{(b)}_{d i'}-{\bar G}_{E_d}(R) J^{(b)}_{di'} {\bigg )}, \ \ R>R_0.
\end{align}
Note that we are treating an energy range thus far where dissociative ionization is not possible, so only $r$ or $R$ can reach infinity in any channel.  Eq. (\ref{eq-wavefunction_joined_full}) has omitted the rapidly decaying $Q$-space Whittaker functions and their coefficients $Z^{(b)}_{QP}$ because they will not play a role in the remaining algebraic steps.

\subsection{Combining the overcomplete set of solutions: the final dissociative columns of the reaction matrix}

%
Note that the low $v^+$ vibrational basis functions in the first sum of Eq.~(\ref{eq-wavefunction_joined_full}) have no $b$ superscript, because they are approximately identical in set $P$ which are functions negligible at and near the dissociative boundary $R_0$.  
%

At this point we have constructed $2N_P$ solutions of the Schr\"odinger equation by collecting solutions with both boundary conditions $b$ at $R_0$.
They are characterized in the outer regions $r>r_0$ and $R>R_0$ by constant matrices right-multiplying regular and irregular outer region solutions, each column of which describes one such solution.
The diagonal matrix of regular solutions whose elements are $f_i(r)$ or ${\bar F}_{E_d}(R)$ is collectively denoted ${\cal F}$, while the arranged diagonal elements of irregular solutions $g_i(r)$ or ${\bar G}_{E_d}(R)$ are denoted ${\cal G}$.
The matrix right-multiplying ${\cal F}$ is the $(N_P+N_D)\times 2N_P$ matrix $I$ and the matrix right-multiplying ${\cal G}$ is the $(N_P+N_D)\times 2N_P$ matrix $J$, giving a compact representation of our $2N_P$ solutions obtained to this point:
\begin{equation}
    \label{Overcomplete}
   \Psi= {\cal F}I-{\cal G}J,
\end{equation}
where the $I= \{I^{(\infty)},I^{(0)}\}$ and $J= \{J^{(\infty)},J^{(0)}\}$, and the first $N_P$ rows of $I^{(b)}$ are simply the identity matrix $\delta_{PP}$ with the final $N_D=3$ rows equal to $I^{(b)}_{dP}$.  Similarly the first $N_P$ rows of $J^{(b)}$ are simply the matrix ${\mathcal K}^{(b)}_{PP}$, with the final $N_D$ rows equal to $J^{(b)}_{dP}$.  It is convenient at this point to combine the set of open and weakly-closed ionization channels $P$ with the $N_D$ dissociation channels $d$, giving $N_P+N_D$ channels represented collectively by the set ${\mathcal P}$.
In fact this combination of the two sets of solutions is overcomplete, because we are constructing ultimately a square solution matrix, creating in the process a square $(N_P+N_D)\times(N_P+N_D)=N_{\mathcal P}\times N_{\mathcal P} $ reaction matrix ${\mathcal K_{{\mathcal P}{\mathcal P}}}$ which is our main goal.  This is accomplished by taking maximally linearly-independent combinations of this combined set of $2N_P$ solutions (i.e., represented by $2N_P$ columns of the above solution matrix in Eq.~(\ref{Overcomplete}), using the Moore-Penrose inverse (or pseudo-inverse) of the enlarged matrix $J$.  This starts by carrying out a singular value decomposition of the matrix $J$, i.e. writing it as 
\begin{equation}
    \label{eq-SVD_of_J}
    J = \mathcal{U} \; \Sigma \; {\tilde{\mathcal{V}}} ,
\end{equation}
where $\mathcal{U}$ is an $(N_P+N_D)\times(N_P+N_D)$ square unitary matrix (here real and orthogonal), $\mathcal{V}$ is a $(2N_P)\times(2N_P)$ square unitary matrix (also real and orthogonal) and $\Sigma$ is an $(N_P+N_D)\times(2N_P)$ rectangular diagonal matrix, also real.  The next step is to right-multiply the overcomplete solution matrix of Eq.~(\ref{Overcomplete}) by $\mathcal{V} \Sigma^{-1}$, and assuming that the number of nonzero singular values equals the expected $N_P+N_D$, the resulting big intermediate reaction matrix with both ionization and dissociation channels, of dimension $N_P+N_D$, has its inverse given by:
\begin{equation}
    \label{eq-K-intermediate}
    {\cal K}^{-1}= I\mathcal{V}\Sigma^{-1}{\tilde{\mathcal{U}}}, 
\end{equation}
and then ${\cal K}$ is found by inversion of ${\cal K}^{-1}$.  This corresponds to the following operation that right-multiplies the solutions represented by Eq.~(\ref{Overcomplete}) by the matrix $\mathcal{V}\Sigma^{-1}{\tilde {\mathcal U}}{\cal K}$.
And of course that same set of transformations applied to the solution matrices must also be applied to the vectors ${\cal D}^{(b)}$ of electric dipole moment transition elements, i.e. 
\begin{equation}
    \label{eq-D_combination}
    {\cal D}={\cal D}^{\rm combined}\mathcal{V}\Sigma^{-1}{\tilde {\mathcal U}}{\cal K},
\end{equation}
where
\begin{equation}
    \label{eq-Dmatrix_rectangle}
    \mathcal{D}^{\rm combined} =\left(
    \begin{array}{c}
        \mathcal{D}^{(\infty)}_{P}  \; , \; \mathcal{D}^{(0)}_{P}
    \end{array}
    \right),
\end{equation}
and this ${\cal D}$ from Eq. (\ref{eq-D_combination}) is then fed into Eq. (\ref{eq-D_final_elim}).
\end{document}